\documentstyle[preprint,aps]{revtex}

\newcommand{\gvz}{\gamma_{vz}}

\newcommand{\gvy}{\gamma_{vy}}
\newcommand{\gve}{\gamma_\varepsilon}

\newcommand{\omb}{\omega_B}

\newcommand{\Bsc}{{\cal B}}

\begin{document}
\draft
\title{Absolute Negative Conductivity and Spontaneous Current Generation 
in Semiconductor Superlattices with Hot Electrons}

\author{Ethan H. Cannon$^{1*}$,
Feodor V. Kusmartsev$^{2}$,
Kirill N. Alekseev$^{3}$,
and David K. Campbell$^1$ }
\address{
$^1$Department of Physics,
University of Illinois at Urbana-Champaign,
1110 West Green St.,
Urbana, IL 61801\\
$^2$Department of Physics,
      Loughborough University,
      Loughborough
      LE11 3TU, UK\\
$^3$Theory of Nonlinear Processes Laboratory,
Kirensky Institute of Physics,
Krasnoyarsk 660036, Russia\\
$^*$ Current address: Department of Electrical Engineering, 
University of Notre Dame,
Notre Dame, IN 46556}

\maketitle
\begin{abstract}
We study electron transport through a semiconductor superlattice 
subject to an electric field parallel to and
a magnetic field perpendicular to the growth axis.
Using a single miniband, semiclassical balance equation model 
with both elastic and inelastic scattering,
we find that (1) the current-voltage characteristic becomes multistable
in a large magnetic field;
and (2) ``hot'' electrons display novel features in their
current-voltage characteristics, including absolute negative conductivity (ANC)
and, for sufficiently strong magnetic fields, a spontaneous dc current
at zero bias.
We discuss possible experimental situations 
providing the necessary hot electrons to observe the predicted ANC
and spontaneous dc current generation.
\end{abstract}
\pacs{PACS numbers: 73.20Dx,  72.80.Ey, 73.40 Gk}

As first realized by Esaki and Tsu \cite{esaki70},
semiconductor superlattices (SSLs) are excellent systems 
for exploring nonlinear transport effects, since
their long spatial periodicity implies that
SSLs have small Brillouin zones and very narrow ``minibands.''
Applied fields accelerate Bloch electrons in a band according to Bloch's
acceleration theorem,
$\dot{\mathbf{k}}=-(e/\hbar)[\mathbf{E}+(\mathbf{v}\times\mathbf{B})/c]$,
where $\mathbf{k}$ is the crystal momentum of the electron, $-e$ its charge, 
$\mathbf{E}$ the electric field, $\mathbf{B}$ the magnetic field, 
$\mathbf{v}$ the electron's velocity, and $c$ the speed of light.
In SSLs, both the velocity and effective mass
depend on the crystal momentum; 
in fact, the effective mass is negative above the band's inflection point,
corresponding to the fact that electrons slow down to zero velocity
as the reach the edge of the Brillouin zone.
The acceleration of the external fields is balanced by 
scattering processes that limit the crystal momentum
of electrons. In clean SSLs with only modest fields,
electrons can reach the negative effective mass (NEM) 
portion of the miniband before scattering.
For an electric field oriented along the SSL growth axis, 
the current-voltage characteristic exhibits a peak followed by 
negative differential conductivity (NDC)
when a significant fraction of electrons explore the NEM region of the miniband
\cite{esaki70};
with an additional magnetic field perpendicular to the growth axis,
NDC occurs at a {\it larger} bias because the magnetic field impedes the 
increase of crystal momentum along the growth axis \cite{ivexpt}.

In this letter, we study electron transport through a single miniband 
of a spatially homogeneous SSL
with growth axis in the $z$-direction in the presence of 
a constant magnetic field, $B$, in the $x$-direction and 
an electric field, $E$, in the $z$-direction.
We assume a tight-binding dispersion relation for the SSL miniband,
$\epsilon({\mathbf{k}})=\hbar^2k_y^2/2m^*+\Delta/2[1-\cos(k_z a)]$,
where $\epsilon$ is the energy of an electron with
crystal momentum $\mathbf{k}$, $m^*$ is the effective mass within the 
plane of the quantum wells (QWs) that form the SSL,
$\Delta$ is the miniband width, and $a$ the SSL period.

Generalizing the approach of \cite{bal_eqns}
to include the effects of the magnetic field, we obtain the following
balance equations \cite{future}
\begin{eqnarray}
\dot{V_y}&=&-\frac{eB}{m^*c}V_z-\gvy V_y \label{eq:be2} \\
\dot{V_z}&=&-\frac{e}{m(\varepsilon_z)}[E-\frac{B V_y}{c}]-\gvz V_z
\label{eq:be3} \\
\dot{\varepsilon_z}&=&-eEV_z+\frac{eB}{c}V_y
V_z-\gve[\varepsilon_z-\varepsilon_{eq,z}]. \label{eq:be4}
\end{eqnarray}
The average electron velocity, ${\mathbf{V}}=(V_y,V_z)$,
is obtained by integrating
the distribution function satisfying the Boltzmann transport equation over
the Brillouin zone; and $\gvy$ and $\gvz$ are the relaxation rates for
the corresponding components of $\mathbf{V}$ following from elastic 
impurity, interface roughness and disorder scattering,
and inelastic phonon scattering.
It is convenient to separate the total energy of the electrons
into parts associated with longitudinal and transverse motion.
Doing so,
$\varepsilon_z$ is the average energy of motion along 
the growth axis with equilibrium value $\varepsilon_{eq,z}$;
$\gve$ represents its relaxation rate due mainly to inelastic phonon
scattering
(elastic scattering that reduces the energy of motion along the 
superlattice growth axis and increases the energy of (transverse)
motion within the QWs
also contributes).
Note that the balance equations contain an effective mass term dependent
on $\varepsilon_z$,
$m(\varepsilon_z)=m_0/(1-2\varepsilon_z/\Delta)$,
which follows from the crystal momentum dependence of the effective mass
tensor;
in this expression, $m_0=2\hbar^2/\Delta a^2$ is the effective mass at the 
bottom of the SSL miniband.
Because of the constant effective mass for motion within the 
plane of the QWs, the energy of this motion does not enter the
balance equations.
While the magnetic field does not change the total electron energy, 
it does transfer energy between in-plane motion and $\varepsilon_z$,
hence Eq. (\ref{eq:be4}) contains the magnetic field-dependent term.

For an intuitive understanding of the balance equations, 
we can consider them as describing an ``average" electron
whose velocity changes according to Newton's second law,
$\dot{\mathbf{V}}={\mathbf{F}}/{\mathbf{m(\varepsilon)}}$,
with $\mathbf{F}$ representing electric, magnetic and damping forces.
The mass tensor ${\mathbf{m(\varepsilon)}}$ is diagonal and 
$m_{zz}$ depends on the energy of motion 
in the $z$-direction; this component of the energy evolves according to
$\dot{\varepsilon_z}=F_z V_z-P_{damp}$.
Inelastic scattering to the average energy $\varepsilon_{eq,z}$ 
(which may not be the bottom of the miniband) 
leads to the damping term, $P_{damp}$.
This gratifyingly intuitive picture should not obscure the
result that our balance equations have been {\it derived} systematically from
the full Boltzmann transport equation.

For numerical simulations, we introduce the scalings
$v_y=((m_0m^*)^{1/2}a/\hbar)V_y$, 
$v_z=(m_0 a/\hbar)V_z$,
$w=(\varepsilon_z-\Delta/2)/(\Delta/2)$,
$w_0=(\varepsilon_{eq,z}-\Delta/2)/(\Delta/2)$,
$\Bsc=eB/(m^*m_0)^{1/2}c$ and
$\omb=eEa/\hbar$ (the Bloch frequency of the electric field).
Note that the average electron energy is scaled such that -1 (+1)
corresponds to the bottom (top) of the miniband.
In terms of the scaled variables, the balance equations read
\begin{eqnarray}
\dot{v_y}&=&-\Bsc v_z-\gvy v_y \label{eq:sbe2} \\
\dot{v_z}&=&\omb w-\Bsc v_yw-\gvz v_z \label{eq:sbe3} \\
\dot{w}&=&-\omb v_z+\Bsc v_yv_z-\gve(w-w_0) \label{eq:sbe4}.
\end{eqnarray}

The current across the superlattice
$I=-eNA(\Delta a/2\hbar)v_{z,ss}$, 
where $N$ is the carrier concentration, $A$ the cross-sectional area and 
$v_{z,ss}$ the steady-state solution to Eq. (\ref{eq:sbe3}).
By setting the time derivatives in Eqs. (\ref{eq:sbe2})-(\ref{eq:sbe4}) to zero,
we obtain the following equation relating
$v_{z,ss}$ and hence the SSL current to the applied bias,
\begin{equation}
C^2v_{z,ss}^3+2C\omb v_{z,ss}^2+[\gvz\gve+\omb^2-\gve w_0C]v_{z,ss}-\gve w_0\omb=0,
\label{eq:IV}
\end{equation}
where $C=\Bsc^2/\gvy$.
This cubic equation for $v_{z,ss}$ implies that there may be up to
three steady-state current values 
for a given bias \cite{epshtein}.

In figure 1, we plot $-v_{z,ss}$, which is proportional to the current across
the SSL, as a function of scaled voltage, $\omb$, 
for various scaled magnetic fields strengths, $C$, 
with equal momentum and energy relaxation rates, $\gvz=\gve$.
With no magnetic field (Fig. 1a), 
the current exhibits a peak followed by negative differential conductance (NDC)
and satisfies the well-known expression 
$v_{z,ss}=(-w_0/\gvz)\omb/(1+\omb^2/\gvz\gve)$
\cite{ktitorov72}.
A magnetic field increases the value of the electric field at which the 
current reaches
its maximum value (Fig. 1b), as has been observed in recent experiments
\cite{ivexpt}.
Finally, for larger magnetic fields (Fig. 1c), 
the current-voltage characteristic from the balance equations 
has a region of multistability with three possible currents.
For SSL parameters of $\Delta=23$ meV, $a=84$\AA, and 
$\gvy=\gvz=\gve=1.5\times 10^{13}$sec$^{-1}$, \cite{ivexpt},
multistability requires a magnetic field of 21 T,
but the semiclassical balance equation is not applicable at such large fields.
However, for SSL parameters of $\Delta=22$meV, $a=90$\AA, and
$\gvy=\gvz=\gve=10^{12}$sec$^{-1}$, \cite{rauch98}, 
multistability should occur for a modest magnetic field of 1.4 T.

Let us now consider the situation of ``hot'' electrons.
In this case the electron distribution is highly non-thermal,
even without the applied fields.
The electrons do not have time to relax to the 
bottom of the miniband before leaving the SSL.
We can effectively describe these hot carriers as relaxing
to the top half of the miniband, {\it i.e,} as having $w_0>0$.
This may happen in a very clean SSL, at very low temperatures,
when the inelastic mean free path is comparable with the SSL size.
The hot electrons may be obtained by injection \cite{rauch97,rauch98,rauch99}
or by an optical excitation.
Below we will discuss how to achieve this situation experimentally.
For zero or small magnetic fields (Fig. 2a), absolute negative conductance (ANC)
occurs as the current flows in the opposite direction as the applied bias.
Then, in larger magnetic fields (Fig. 2b), a region of multistability appears
around zero bias;
a linear stability analysis shows the zero current solution becomes unstable
as soon as the nonzero current solutions emerge. 
{\it In other words, the SSL will spontaneously develop a current across it
at zero bias.}
The three possible steady-state velocities at zero bias are
\begin{equation}
v_{z,ss}=0,\pm(\frac{\gve w_0 C-\gvz\gve}{C^2})^{1/2},
\label{eq:Izb}
\end{equation}
so a spontaneous current will appear when 
the condition $w_0C>\gvz$ (in other words, $w_0\Bsc^2/\gvy>\gvz$) 
is satisfied.
Since $C$ and $\gvz$ are always positive, this requires that $w_0$ be positive;
neither thermal effects nor doping can fulfill 
the necessary condition for a 
zero-bias current: hot electrons are required. 
Physically, one clearly needs energy to create the spontaneous current, and
this energy is supplied by hot electrons.

These two results for hot electrons, 
{\it i.e.} ANC and spontaneous current generation, 
follow from their negative effective mass in the top half of the miniband. 
To understand the origin of the ANC, consider a one-dimensional SSL 
with electrons at their equilibrium position at the bottom of the band,
$w_0=-1$, and no electric field;
when a positive bias is applied, $\omb>0$, the electrons move through
the band according to $\dot{k_za}=-\omb$ until a scattering event occurs.
Elastic scattering conserves energy, sending an electron across the band
in this one-dimensional case.
Inelastic scattering changes the electron energy to $w_0$,
{\it i.e.} $k_z$=0.
(Hot electrons inelastically scatter to $w_0>0$, possibly gaining energy.)
In Fig. 3, the electric field accelerates the electrons from their
equilibrium position at point A; inelastic scattering prevents many electrons
from passing point B, so electrons are found mainly in the segment AB.
Elastic scattering sends electrons into the segment AC, which contains fewer
electrons than the segment AB.
In this tight-binding miniband, the velocity of an electron with
crystal momentum $k_z$ is 
${\cal V}(k_z)\equiv\hbar^{-1}\partial\epsilon/\partial k_z=(\Delta a/2\hbar)\sin(k_za)$;
because the segment AB has more electrons, there is a net negative velocity,
or a positive current, as expected for a positive voltage.
In the presence of hot electrons,
when $w_0>0$, the two points labeled D1 and D2 initially are occupied 
with equal numbers of electrons and no current flows.
Once applied, the electric field accelerates electrons such that they
occupy the segments D1E1 and D2E2, as inelastic (non-energy-conserving) 
scattering returns them to their quasi-equilibrium energy at points D1 and D2; 
elastic scattering leads to a smaller number of electrons in 
the segments D1F1 and D2F2.
The speed of electrons above the inflection point of the miniband decreases
as the magnitude of their crystal momentum increases towards the edge of
the Brillouin zone,
thus the electrons in the segment D2E2 have a larger speed than those in the
segment D1E1. 
A positive net velocity or, in other words, a negative current results;
this is the absolute negative conductivity shown in figure 2a.

An intuitive picture of the spontaneous current generation also follows
from the miniband structure of an SSL in an external magnetic field.
Consider a small, positive current fluctuation across the SSL 
resulting from extra electrons at the initial energy $w_0>0$, 
point D2 in Fig. 3.
The crystal momentum evolves according to 
$\dot{\mathbf{k}}=-(e/\hbar c)\mathbf{\cal V}\times\mathbf{B}$;
with $B_x>0$, initially $\dot{k_y}>0$, hence $\dot{k_z}<0$.
The electron moves from point D2 towards E2 with increasing speed,
until inelastic scattering returns the electron to its 
quasi-equilibrium initial position or
elastic scattering sends it across the band.
For a large enough magnetic field, small enough elastic scattering and 
electrons far enough into the NEM region of the miniband
(as specified by the requirement $w_0\Bsc^2>\gvy\gvz$),
the initial current fluctuation will increase, 
the zero current state will be unstable to such small fluctuations,
and the SSL will develop a spontaneous current.

Experimentally, it is possible to obtain these hot electrons with $w_0>0$
by injecting electrons into the NEM portion of the miniband, 
as was described recently in references \cite{rauch97,rauch98,rauch99}.
In this injection geometry, two mechanisms contribute 
to the current through the SSL:
first, coherent tunneling through the whole SSL, and, second,
incoherent transport of scattered electrons that do not maintain phase 
information \cite{rauch98}.
The balance equations describe these latter electrons.
While the electrons in the NEM region can support a current instability, 
those that have scattered to the bottom of the miniband cannot,
so it is vitally important to keep the miniband width below the LO phonon 
energy of 36 meV in order to limit phonon scattering.
In this case, the balance equations describe the behavior of
electrons that have scattered elastically, primarily because of disorder.
When the injection energy is in the forbidden region
below the miniband, there is no appreciable
current through the SSL;
as the injection energy is swept through the miniband, the current increases, 
since electrons incident at the miniband energy can traverse the SSL.
The current then decreases again when the injection energy passes through 
the top half of the miniband.
The sharpness of this decrease depends on the miniband width
because phonon replicas emerge when electrons having undergone LO phonon
scattering are at the miniband position.
For a sharp feature, a narrow miniband is important
(see Fig. 3 in reference \cite{rauch97}),
such that the width of the incident wavepacket (about 17meV)
plus the miniband width is less than the LO phonon energy (36meV).

To observe the hot electron effects we predict, the transmitted current-
injection energy characteristic must first be measured with no external fields;
the experiment must then be repeated with a magnetic field in the plane of
the quantum well.
While such a field reduces the coherent current \cite{rauch99}, 
if sufficiently strong, it can lead to spontaneous current generation for
electrons incident near the top of the miniband, 
{\it i.e.} between the peaks in the current-injection energy curve.
This current instability would cause the current to flatten, or even
increase, between the main peak and its phonon replica.
To observe the ANC, the current-injection energy curves must be measured for 
positive and negative voltages;
it is known that in a positive or negative bias 
the location of the current peak shifts due to the voltage drop across the 
drift region and the coherent current decreases \cite{rauch98}.
When the phonon replica current is small, it may also be possible to observe
a change in the shape of the current peak.
For positive bias, the current below the peak, at injection energies in the
lower half of the miniband, increases.
Meanwhile the current above the peak, for injection energy in the top
half of the miniband, decreases due to ANC; 
the current drops off more rapidly on the high-injection energy side of the
current peak.
Just the opposite occurs for a negative bias: since ANC causes the current
to increase for injection energies in the top half of the miniband, 
the peak drops less sharply.
Finally, we note that recently Kempa and coworkers have studied 
the possibility of generating non-equilibrium plasma instabilities 
through a similar selective energy injection scheme \cite{kempa}.
The other possibility to create the hot electrons is
an optical excitation of electron-hole pairs.
As far as we aware, this approach has not yet been used
specifically as a method
of injecting hot electrons into an SSL. 

In summary, we have described new physical effects----incoherent
current flow opposite to the direction of the applied electric bias
and spontaneous current generation for hot electrons in
a transverse magnetic field---in an SSL with
nonequilibrium electron excitations and
have suggested how they might be observed in experiments.
We hope our experimental colleagues will search for
these effects.

We are grateful to Lawrence Eaves for stimulating discussions.
F.V.K. thanks the Department of Physics at the University of Illinois at
Urbana-Champaign for its hospitality.
This work was partially supported by NATO Linkage Grant NATO LG 931602 and 
INTAS.
E.H.C. acknowledges support by a graduate traineeship under NSF GER93-54978.

\begin{figure}
\caption{Scaled current-voltage characteristic for an SSL 
with equal momentum and energy relaxation rates, $\gvz=\gve$, 
at zero temperature, $w_0=-1$, with no magnetic field, $C=0$ (a), 
in a small magnetic field, $C=\gvz$ (b), 
and in a large magnetic field, $C=15\gvz$ (c).}
\label{fig1}
\end{figure}
\begin{figure}
\caption{Scaled current-voltage characteristic for an SSL
with equal momentum and energy relaxation rates and hot electrons,
$w_0=0.5$, in a small magnetic field, $C=\gvz$ (a), 
and in a large magnetic field, $C=5\gvz$ (b).}
\label{fig2}
\end{figure}
\begin{figure}
\caption{Miniband structure for SSL: scaled energy vs. 
crystal momentum along growth axis.  Labeled points are described in text.}
\label{fig3}
\end{figure}

\begin{references}
%
\bibitem{esaki70}
L.~Esaki and R.~Tsu, IBM J. Res. Dev. {\bf 14}, 61 (1970).
%
\bibitem{ivexpt}
F.~Aristone {\it et al.}, Appl. Phys. Lett. {\bf 67}, 2916 (1995);
L.~Canali {\it et al.}, Superlattices Microstruct. {\bf 22}, 155 (1997).
%
\bibitem{bal_eqns}
A.~A. Ignatov and V.~I. Shashkin, Phys. Lett. A {\bf94}, 169 (1983).
%
\bibitem{future}
E. H. Cannon, K. N. Alekseev, D. K. Campbell, F. V. Kusmartsev, unpublished.
%
\bibitem{epshtein}
{\'{E}}.~M. {\'{E}}pshtein, Radiophysics and Quantum Electronics
(Consultant's Bureau) {\bf22}, 259 (1979), 
Sov. Phys. Semicond. {\bf 25}, 216 (1991). In these
references, {\'E}pshtein discussed a
similar effect---namely, the Hall field across a current-biased superlattice 
to lowest order in the magnetic field---and found that
the Hall voltage becomes multivalued for sufficient current; 
in contrast, we discuss the current across a voltage-biased superlattice 
in a magnetic field of arbitrary strength (although our semiclassical
model breaks down in a quantizing magnetic field).
%
\bibitem{ktitorov72}
S.~A. Ktitorov, G.~S. Simin, and V.~Ya. Sindalovskii, Sov. Phys. Solid State
{\bf 13}, 1872 (1972);
A.~A. Ignatov, E.~P. Dodin, and V.~I. Shashkin, Mod. Phys. Lett. B {\bf 5},
1087 (1991).
%
\bibitem{rauch98}
C.~Rauch {\it et al.}, Phys. Rev. Lett. {\bf 81}, 3495 (1998).
%
\bibitem{rauch97}
C.~Rauch {\it et al.}, Appl. Phys. Lett. {\bf 70}, 649 (1997).
%
\bibitem{rauch99}
C.~Rauch {\it et al.}, Superlattices Microstruct. {\bf 25}, 47 (1999).
%
\bibitem{kempa}
K.~Kempa, P.~Bakshi, and E.~Gornik, Phys. Rev. B {\bf 54}, 8231 (1996);
K.~Kempa {\it et al.}, J. Appl. Phys. {\bf 85}, 3708 (1999).

\end{references}
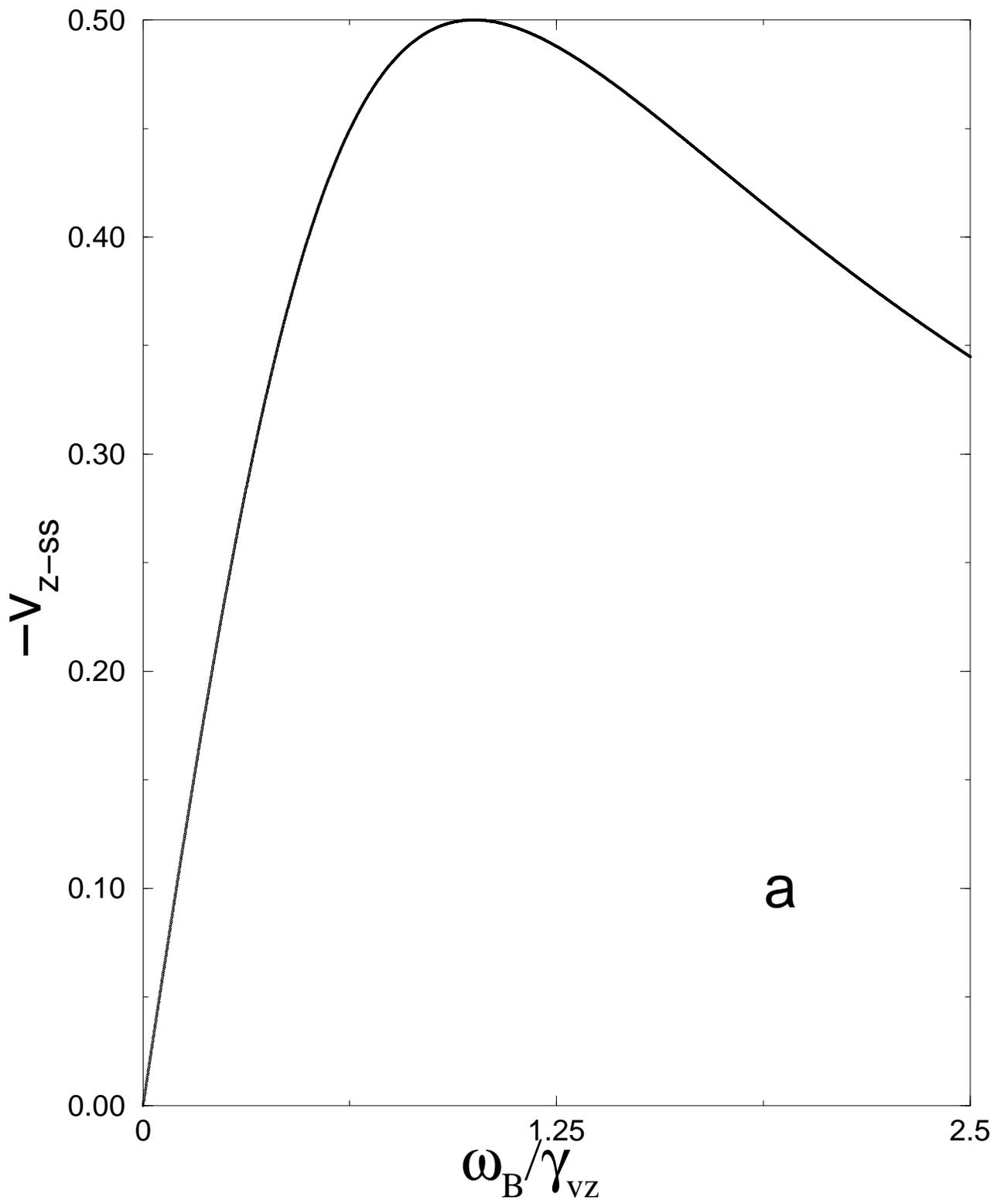
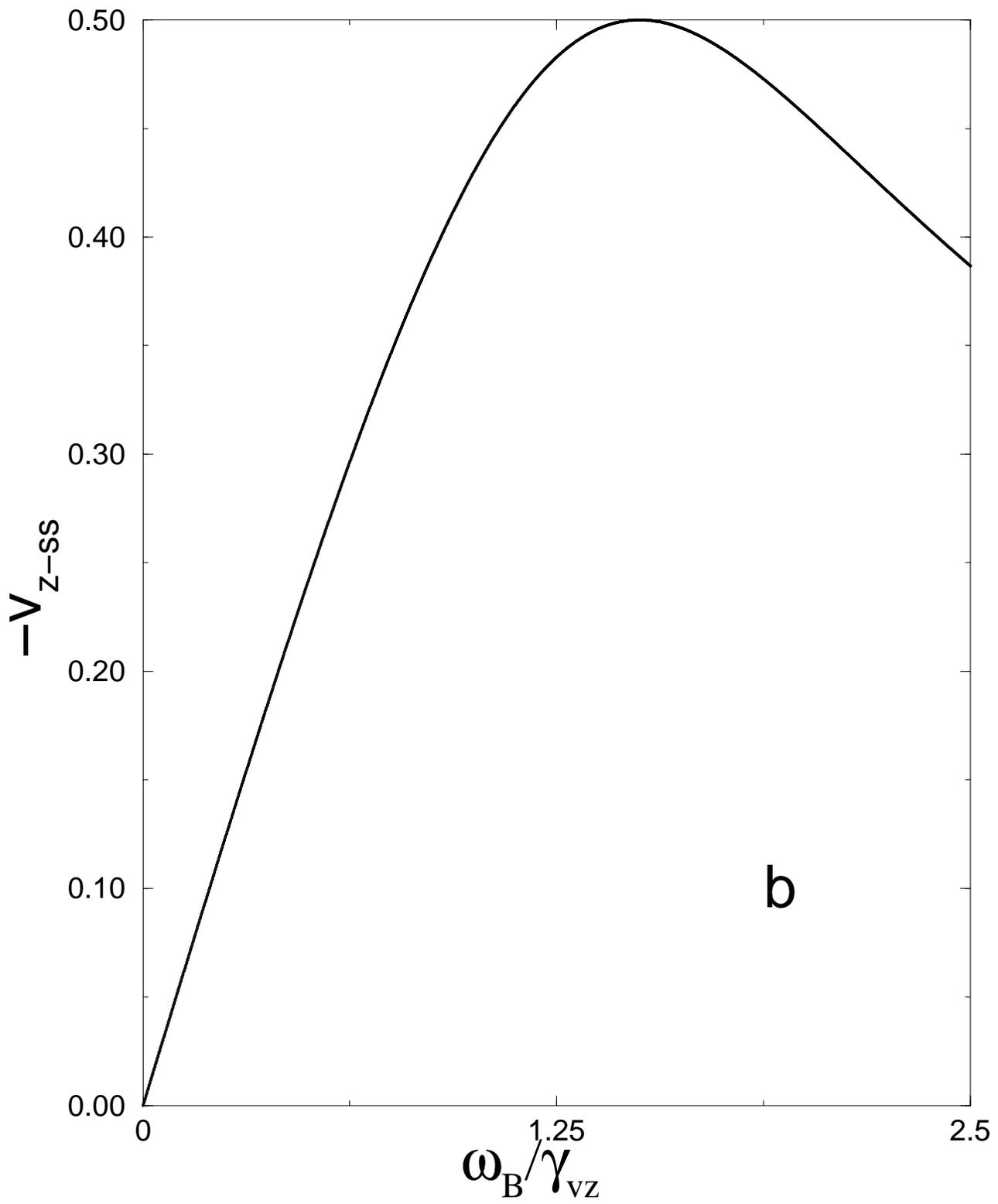
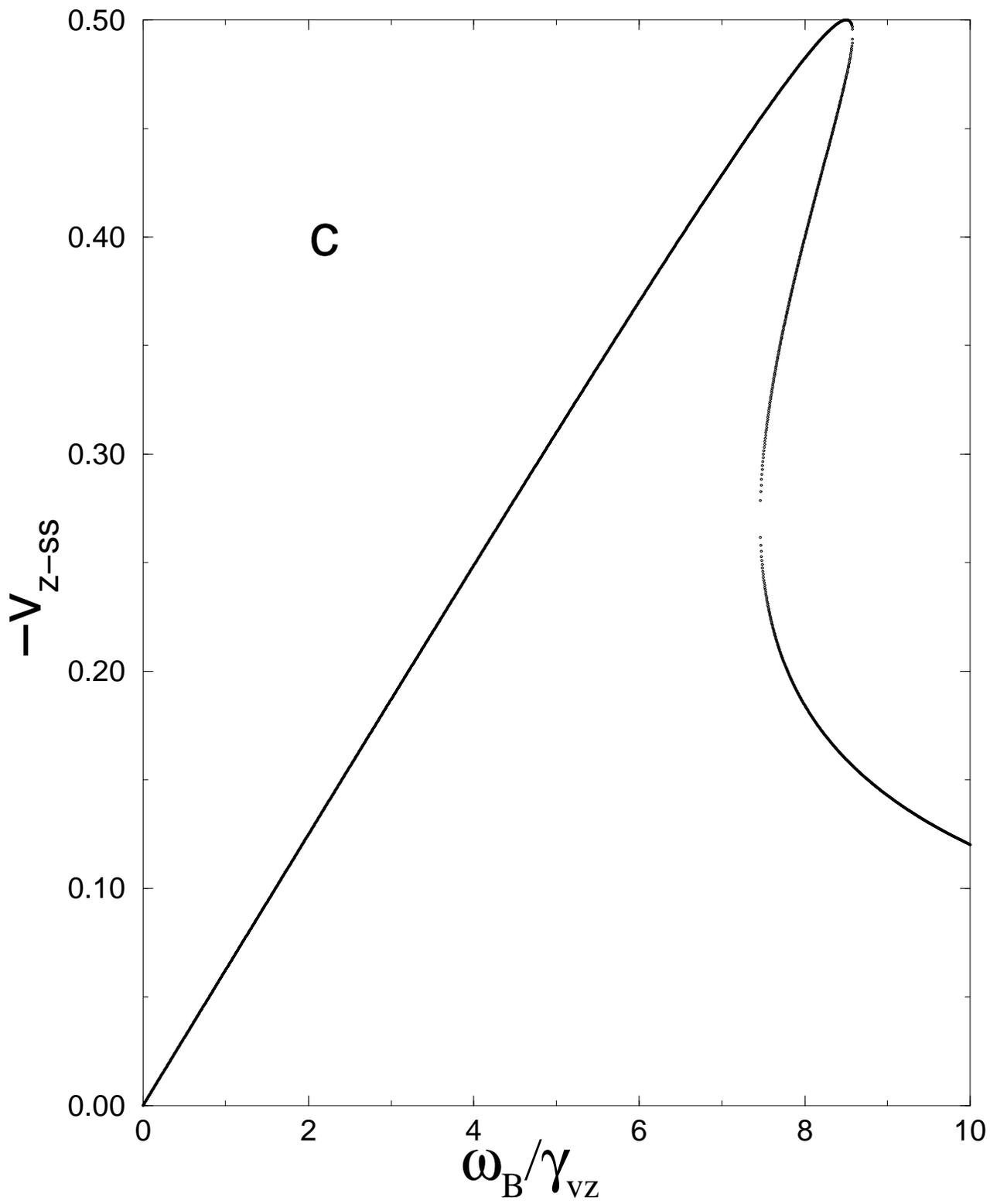
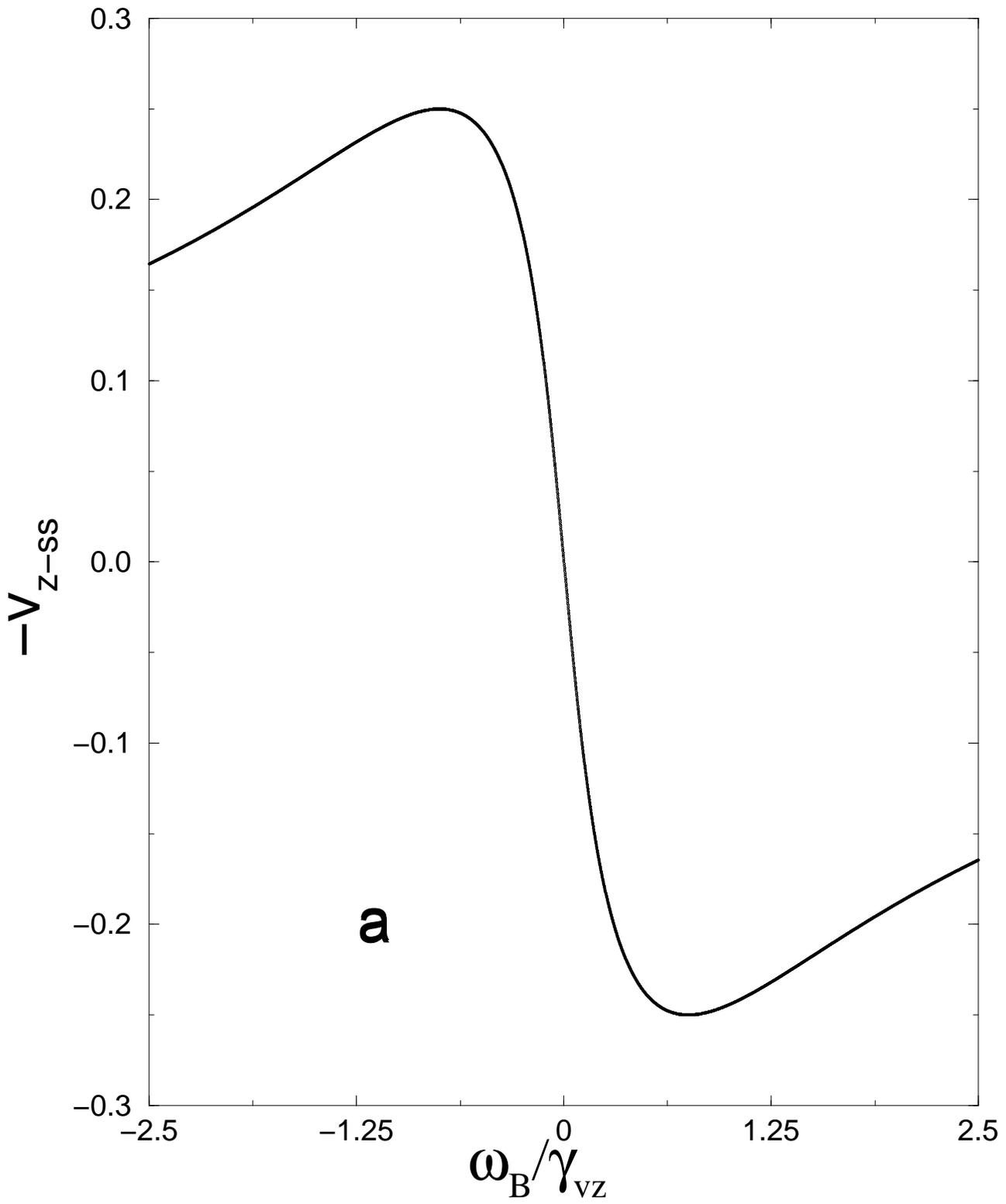
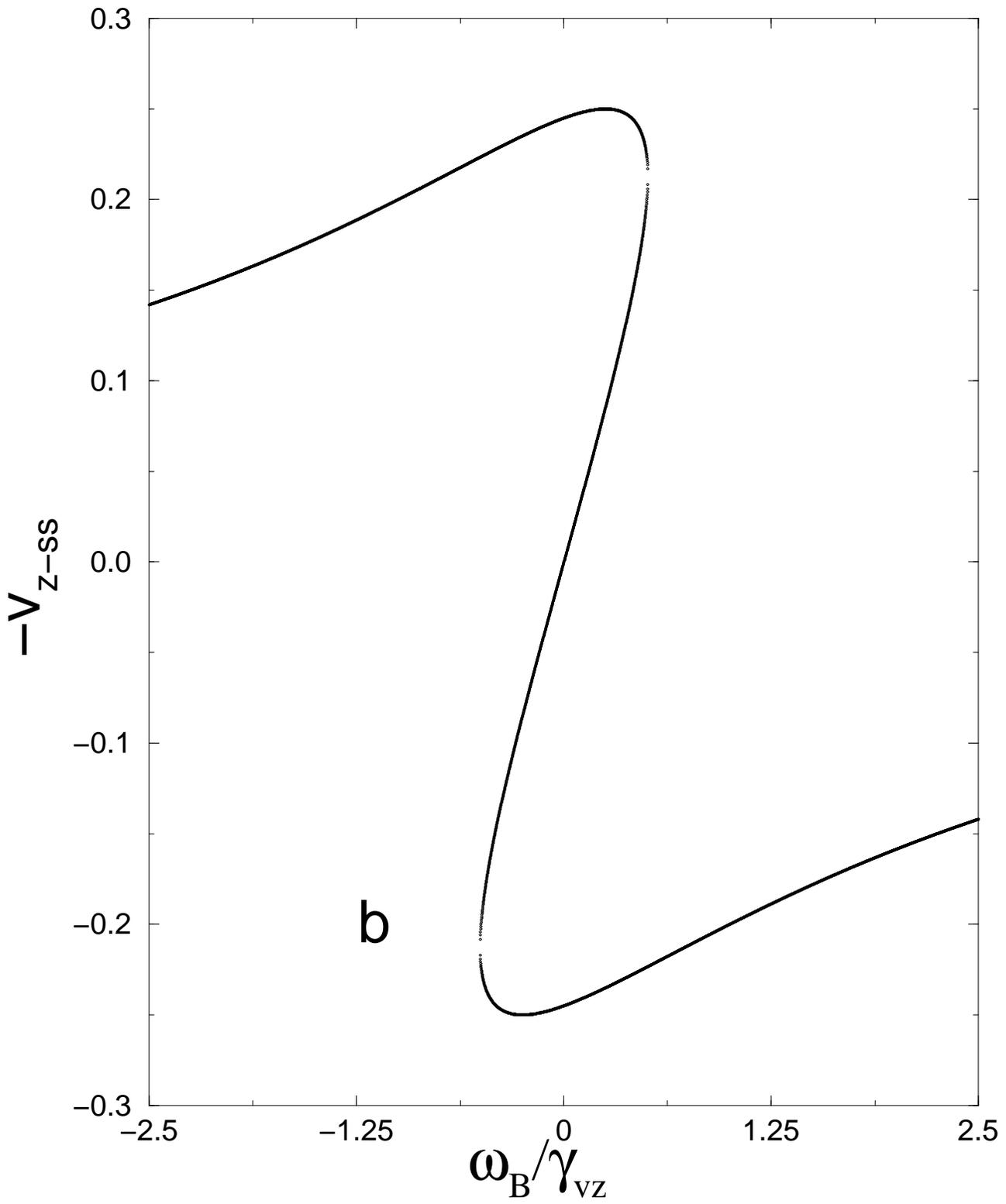
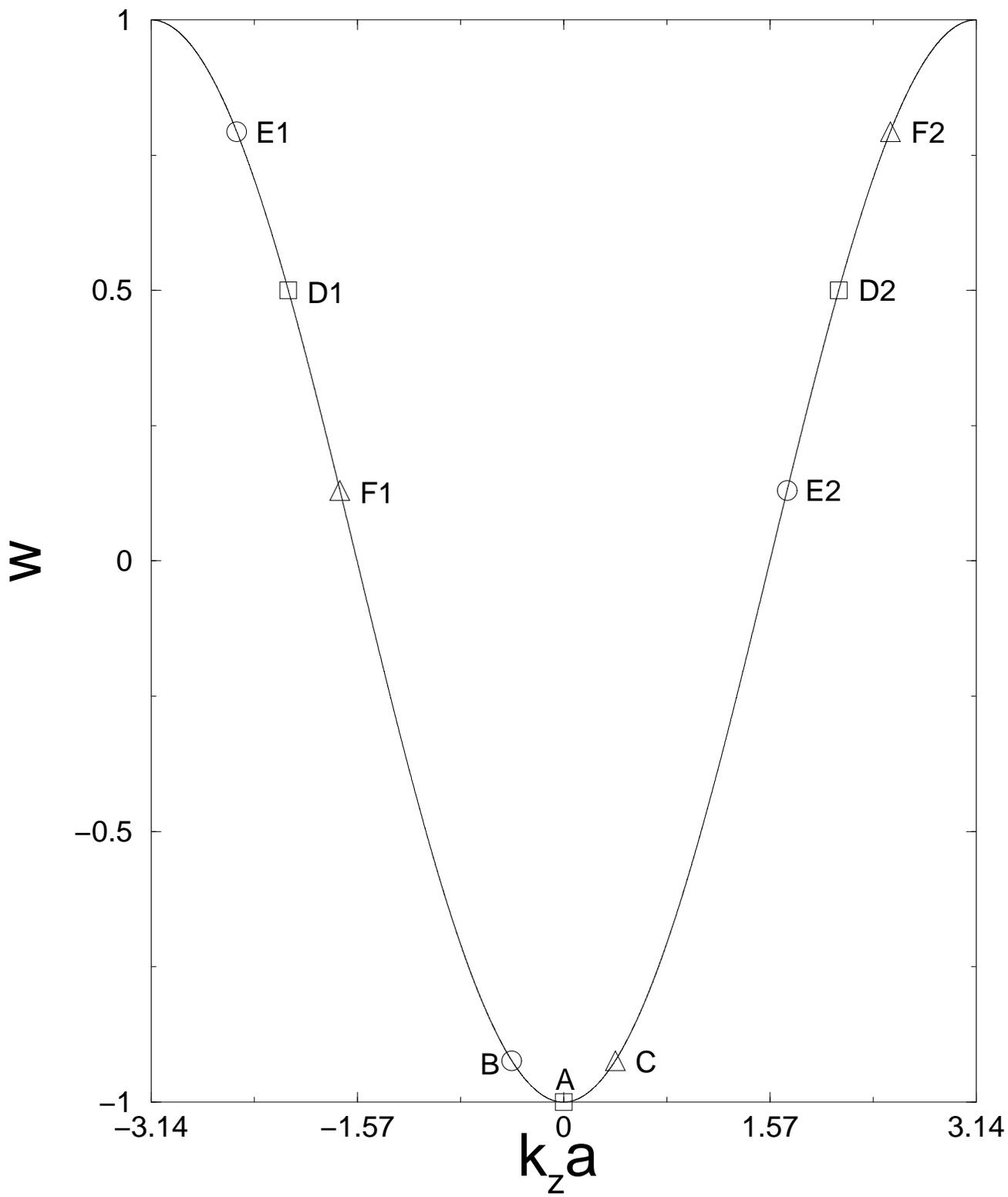
\end{document}